\documentstyle[12pt,epsf,epsfig,a4]{article}
\begin{document}
\begin {titlepage}
\begin{flushleft}
FSUJ TPI QO-14/98
\end{flushleft}
\begin{flushright}
November, 1998
\end{flushright}
\vspace{20mm}

\begin{center}
{\Large \bf Conditional quantum-state transformation at a beam splitter } 
\vspace{15mm}
 
{\large \bf J. Clausen, M. Dakna, L. Kn\"oll and D.--G. Welsch}\\[1.ex]
{\large Friedrich-Schiller-Universit\"at Jena}\\[0.5 ex]
{\large Theoretisch-Physikalisches Institut}\\[0.5 ex]
{\large Max-Wien Platz 1, D-07743 Jena, Germany}
\vspace{25mm}
\end{center}
\begin{center}
\bf{Abstract}
\end{center}
Using conditional measurement on a beam splitter, 
we study the transformation of the quantum state
of the signal mode within the concept of two-port non-unitary 
transformation. Allowing for arbitrary quantum 
states of both the input reference mode and the output 
reference mode on which the measurement is performed, 
we show that the non-unitary transformation operator
can be given in a closed form by an $s$-ordered operator product,
where the value of $s$ is entirely determined by the 
absolute value of the beam splitter reflectance (or transmittance). 
The formalism generalizes previously 
obtained results that can be recovered by simple
specification of the non-unitary transformation operator.
As an application, we consider the generation 
of Schr\"odinger-cat-like states. An extension to
mixed states and imperfect detection is outlined.
\end{titlepage}
\section{Introduction}
\label{1}
Entanglement is one of the most striking features of quantum mechanics. 
Roughly speaking, a quantum state of a system composed of subsystems is 
said to be entangled, if it cannot be decomposed into a product of states 
of the subsystems and the correlation is nonclassical (note that there 
is no generally accepted definition of the degree of entanglement 
\cite{Schlienz}). Recently applications of entangled quantum states in 
quantum information processing have been extensively 
discussed \cite{Steane}. Entangled quantum states also offer novel 
possibilities of quantum state engineering using conditional measurement. 
One of two entangled quantum objects is prepared in a desired state owing 
to the state reduction associated with an appropriate measurement 
on the other object. The quantum state of travelling optical modes can 
be entangled, e.g., by mixing the modes at an appropriately chosen 
multiport. The simplest example is the superposition of two modes by a
beam splitter. Combination of beam splitters with measuring instruments 
in certain output channels may therefore be regarded as a promising way 
for engineering quantum states of travelling optical fields 
\cite{Ban1,Dakna1,Dakna2,Dakna3,10,11,12,13,Dakna4}.  

The action of a beam splitter as a lossless four-port device is 
commonly described in terms of a unitary transformation 
connecting the two input fields and the two output fields 
\cite{Campos}. With regard to conditional measurement, it is
convenient to regard the combined action on the signal of the beam 
splitter and the measuring instrument as the action of an optical 
two-port device. Following this concept, a non-unitary transformation 
operator in the Hilbert space of the signal field can be introduced
which is independent of the signal quantum state. 
In this paper we present this operator in a closed form for 
arbitrary input quantum states of the reference mode and
arbitrary quantum states measured in the output channel
of that mode. 

The developed formalism generalizes and unifies previous approaches
to the problem of conditional measurement on a beam splitter 
\cite{Ban1,Dakna1,Dakna2,Dakna3}
and enables us to calculate the conditional output states in a very 
straightforward way. To illustrate the formalism, we consider 
the generation of Schr\"odinger-cat-like states from coherent and 
Fock states and give a brief extension to the generation of
multiple Schr\"odinger-cat-like states.  
Originally introduced for probing the foundations of quantum mechanics, 
quantum superposition states of Schr\"odinger-cat-type 
(for a review, see \cite{Buzek}) have recently been 
suggested to be applied as logical qubits in quantum 
computing \cite{Cochrane}.

The paper is organized as follows. Section \ref{Sec2}
introduces the basic-theoretical concept and presents 
the non-unitary transformation operator. In Section \ref{Sec3}
the formalism is applied to the generation of
Schr\"odinger-cat-like states, and a summary is given
in Section \ref{Sec5}.

\section{The conditional beam splitter operator
\label{Sec2}}

\begin{figure}
\centering\epsfig{figure=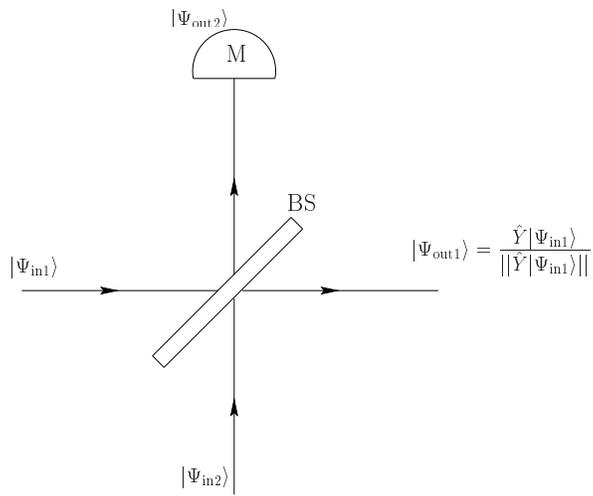,width=0.8\linewidth}
\caption{After mixing a signal mode prepared in the state $|\Psi_{{\rm in}_1}\rangle$ 
with a reference mode prepared in the state $|\Psi_{{\rm in}_2}\rangle$, 
the output signal mode collapses to the state $|\Psi_{{\rm out}_1}\rangle$, 
if the measuring instrument M projects the output reference mode onto the state 
$|\Psi_{{\rm out}_2}\rangle$.
\label{Fig1}}
\end{figure}

Let us consider an experimental setup of the type shown in 
figure~\ref{Fig1}. A signal mode 
(index 1) is mixed with a reference mode (index 2) at a beam splitter,
and a measurement (device M) is performed on the output reference mode.
Since the two output modes are entangled in general, 
the measurement influences the output signal mode as well. The action 
of a beam splitter can be described by a unitary operator $\hat{U}$
connecting the input and output states according to 
\begin{equation}
  |\Psi_{{\rm out}}\rangle = \hat{U}|\Psi_{{\rm in}}\rangle\,,
\label{2.1}
\end{equation}
where \cite{Campos}
\begin{equation}
\hat{U} =  e^{\i(\varphi_T + \varphi_R)\hat{L}_3}
 e^{2i\vartheta\hat{L}_2} e^{\i(\varphi_T -\varphi_R)\hat{L}_3},
\label{2.2}
\end{equation}
with
\begin{equation}
\hat{L}_2=\frac{1}{2\i}(\hat{a}_1^\dagger\hat{a}_2-\hat{a}_2^+\hat{a}_1),\qquad
\hat{L}_3=\frac{1}{2}(\hat{a}_1^\dagger\hat{a}_1-\hat{a}_2^\dagger\hat{a}_2)\,, 
\label{2.3}
\end{equation}
and the complex transmittance $T$ and reflectance $R$ of the beam splitter 
are defined by 
\begin{equation}
T= e^{\i\varphi_T}\cos\vartheta,\qquad R= e^{\i\varphi_R}\sin\vartheta.
\label{2.3a}
\end{equation}
Now let us assume that $\hat{\Pi}(l)$ is the positive operator 
valued measure (POVM) that is realized by the measuring device M, with
\begin{eqnarray}
  \hat{\Pi}(l) \ge 0, \qquad 
  \sum_l\hat{\Pi}(l)=1
\end{eqnarray}
(for POVM, see, e.g., \cite{Helstrom,Busch}). When the measurement 
on the output reference mode yields the result $l$, then
the reduced state of the output signal mode becomes
\begin{equation}
  \hat{\varrho}_{{\rm out}_1} =
 \frac{ {\rm Tr}_2\!\left[\hat{\varrho}_{{\rm out}}\hat{\Pi}(l)\right]}{p(l)}\,,
\qquad \hat{\varrho}_{{\rm out}}= \hat{U}|\Psi_{{\rm in}}\rangle
\langle\Psi_{{\rm in}}|\hat{U}^\dagger,
\label{2.4}
\end{equation}
where
\begin{equation}
p(l) = \langle\hat{\Pi}(l)\rangle = 
{\rm Tr}_1{\rm Tr}_2\!\left[\hat{\varrho}_{{\rm out}}\hat{\Pi}(l)\right]
\label{2.5}
\end{equation}
is the probability of obtaining the result $l$.
In particular, when $\hat{\Pi}(l)$ projects onto a 
pure state $|l\rangle$ $\!=$ $\!|\Psi_{{\rm out}_2}\rangle$, i.e., 
\begin{equation}
\hat{\Pi}(l) 
= |\Psi_{{\rm out}_2}\rangle\langle\Psi_{{\rm out}_2}|,
\label{2.6}
\end{equation}
and the (pure) input state can be decomposed as 
\begin{equation}
|\Psi_{{\rm in}}\rangle = |\Psi_{{\rm in}_1}\rangle|\Psi_{{\rm in}_2}\rangle,
\label{2.7}
\end{equation}
then combination of equations (\ref{2.4}) -- (\ref{2.7}) yields
\begin{equation}
\label{2.8}
\hat{\varrho}_{{\rm out}_1} = 
|\Psi_{{\rm out}_1}\rangle\langle\Psi_{{\rm out}_1}|,
\end{equation}
with
\begin{equation}
|\Psi_{{\rm out}_1}\rangle=\frac{\hat{Y}|\Psi_{{\rm in}_1}\rangle}
{\|\hat{Y}|\Psi_{{\rm in}_1}\rangle\|} \,,
\label{2.9}
\end{equation}  
where  
\begin{equation}
\hat{Y}=\langle\Psi_{{\rm out}_2}|\hat{U}|\Psi_{{\rm in}_2}\rangle,
\label{2.10}
\end{equation}
is the non-unitary conditional beam splitter operator $\hat{Y}$ 
defined in the signal-mode Hilbert space, 
the expectation value of $\hat{Y}^\dagger\hat{Y}$ being the 
probability of obtaining the state $|\Psi_{{\rm out}_1}\rangle$, 
\begin{equation}
\label{2.11}
  p(\Psi_{{\rm out}_1})=\|\hat{Y}|\Psi_{{\rm in}_1}\rangle\|^2
\equiv 
\langle \Psi_{{\rm in}_1}|\hat{Y}^\dagger 
\hat{Y}|\Psi_{{\rm in}_1}\rangle.
\end{equation}

In order to determine the non-unitary transformation operator $\hat{Y}$,
let us first consider reference modes that are prepared in displaced Fock
states (for displaced Fock states, see, e.g., \cite{Nieto} and references 
therein),
\begin{eqnarray}
\label{2.12}
|\Psi_{{\rm in}_2}\rangle = \hat{D}_2(\alpha)|m\rangle_2\label{d1}\,,
\qquad
|\Psi_{{\rm out}_2}\rangle&=&\hat{D}_2(\beta)|n\rangle_2\label{d2}\,,
\end{eqnarray}
with $\hat{D}(\alpha)$ $\!=$ $\!{\rm exp}(\alpha\hat{a}^\dagger$
$\!-$ $\!\alpha^*\hat{a})$ being the coherent displacement operator. 
After a lengthy but straightforward calculation (see \ref{app1})
we find that 
\begin{equation}
  \hat{Y} = \hat{D}_1\!\left(\frac{\alpha-T\beta}{R^\ast}\right)
  \,_2\langle n|\hat{U}|m\rangle_2\,
  \hat{D}_1\!\left(\frac{\beta-T^*\alpha}{R^*}\right),
\label{2.14}
\end{equation}
where
\begin{equation}
  _2\langle n|\hat{U}|m\rangle_2 = \frac{R^m(-R^*)^n}{T^n\sqrt{m!n!}}
  \left\{(\hat{a}_1^\dagger)^m\hat{a}_1^n\right\}_sT^{\hat{n}_1}.
\label{2.15}
\end{equation}
Here, the notation $\{\cdots\}_{s}$ introduces $s$-ordering 
(for $s$-ordering, see \cite{Cahill}), with
\begin{equation}
s=\frac{2}{|R|^2}-1.
\label{2.16}
\end{equation}
Note that $s$ $\!>$ $\!1$. Applying equation (\ref{ordering}) for $t$ $\!=$
$\!1$ and using the formulas given in Appendix A in Ref.~\cite{Dakna3},
the $s$-ordered operator product in equation (\ref{2.15}) can be rewritten as
\begin{eqnarray}
\hspace{-4ex}
\{(\hat{a}^\dagger)^m\hat{a}^n\}_s
\!=\!\left\{
\begin{array}{c@{\quad{\rm if}\quad}l}
\displaystyle
m!\left[-\frac{s\!+\!1}{2}\right]^m\hat{a}^{n-m}
\,{\rm P}_m^{(n-m,\hat{n}-n)}\!\left[\frac{s\!-\!3}{s\!+\!1}\right] 
& m\le n, \\[2.5ex]
\displaystyle
n!\left[-\frac{s\!+\!1}{2}\right]^n(\hat{a}^\dagger)^{m-n}\,
{\rm P}_n^{(m-n,\hat{n}-n)}\!\left[\frac{s\!-\!3}{s\!+\!1}\right] 
& m\ge n,
\end{array}\right.
\label{2.17}
\end{eqnarray}
where ${\rm P}_a^{(b,c)}(z)$ is the Jacobi
polynomial. It should be mentioned that when $\alpha$ $\!=$ $\!\beta$ 
$\!=$ $\!0$, then the operator $\hat{Y}$ in equation (\ref{2.14}) realizes the 
transformation to the Jacobi-Polynomial states  
in Ref.~\cite{Dakna3}.
 
Now, the generalization of the formalism to arbitrary (pure) 
quantum states of the reference modes, 
\begin{eqnarray}
\label{2.18}
|\Psi_{{\rm in}_2}\rangle=\hat{F}(\hat{a}_2^\dagger)|0\rangle_2\,,
\qquad
|\Psi_{{\rm out}_2}\rangle=\hat{G}(\hat{a}_2^\dagger)|0\rangle_2\,,
\end{eqnarray}
is straightforward. From equations (\ref{2.10}) and (\ref{2.18}) and 
application of the relation (\ref{2.15}) $\hat{Y}$ is obtained to be 
\begin{equation}
\hat{Y} = \left\{\hat{F}(R\hat{a}_1^\dagger)\,
\hat{G}^\dagger\!\left(-\frac{R}{T^*}\hat{a}_1^\dagger\right)
\right\}_sT^{\hat{n}_1}.
\label{2.20}
\end{equation}
The result reveals, that (up to the operator $T^{\hat{n}_1}$) the 
non-unitary conditional beam splitter operator $\hat{Y}$ is nothing
but the operator product $\hat{F}\hat{G}^\dagger$ in $s$ order,
with $s$ from equation (\ref{2.16}).
Although equation (\ref{2.20}) already covers the general case, it may 
be useful, for practical reasons, to consider explicitly 
coherently displaced quantum states of the reference modes, that is
\begin{eqnarray}
\label{2.21}
|\Psi_{{\rm in}_2}\rangle=\hat{D}_2(\alpha)\hat{F}(\hat{a}_2^\dagger)
|0\rangle_2\,,
\qquad
|\Psi_{{\rm out}_2}\rangle=\hat{D}_2(\beta)\hat{G}(\hat{a}_2^\dagger)
|0\rangle_2\,.
\end{eqnarray}
In close analogy to the derivation of equations (\ref{2.14}) and
(\ref{2.20}) we find that  
\begin{equation}
  \hat{Y} = \hat{D}_1\!\left(\frac{\alpha-T\beta}{R^*}\right)\,
  \left\{\hat{F}(R\hat{a}_1^\dagger)\,\hat{G}^\dagger\!\left(-\frac{R}
  {T^*}\hat{a}_1^\dagger\right)\right\}_sT^{\hat{n}_1}\,
  \hat{D}_1\!\left(\frac{\beta-T^*\alpha}{R^*}\right),
\label{2.23}
\end{equation}
which shows that displacing the quantum states of the reference modes
always leads to displaced quantum states of the signal modes.
 
 From equation (\ref{2.23}) it is easily seen that the ordering procedure 
can be omitted if at least one of the reference modes is prepared in a 
coherent state (the vacuum included), i.e., $\hat{F}$ $\!=$ $\!\hat I$ or 
$\hat{G}$ $\!=$ $\!\hat I$. This is the case when, e.g.,  
$|\Psi_{{\rm in}_2}\rangle$ $\!=$ $\!|0\rangle_2$ and 
$|\Psi_{{\rm out}_2}\rangle$ $\!=$ $\!|n\rangle_2$
(or $|\Psi_{{\rm in}_2}\rangle$ $\!=$ $\!|n\rangle_2$ and 
$|\Psi_{{\rm out}_2}\rangle$ $\!=$ $\!|0\rangle_2$), which leads to
preparation of the output signal mode in
a photon-subtracted (or photon-added) state \cite{Dakna1,Dakna2}.
Note that preparation of the output reference mode in a coherent
state can be realized in eight-port homodyne detection
or heterodyne detection.
Further it should be mentioned that $|\Psi_{{\rm in}_1}\rangle$ and 
$|\Psi_{{\rm in}_2}\rangle$ can be interchanged if $T$ and $R$ are 
replaced with i$R$ and i$T$, respectively, because of the symmetry of 
the beam splitter transformation \cite{Campos}. The operator $\hat{Y}$ 
then transforms (up to a global phase factor) the state 
$|\Psi_{{\rm in}_2}\rangle$ into the state $|\Psi_{{\rm out}_1}\rangle$. 
In this way the $s$-ordering procedure can also be circumvented when 
the signal mode is prepared in a coherent state. 

In general, the POVM realized by the 
measurement apparatus does not project onto a pure state, and 
equation (\ref{2.6}) must be replaced with
\begin{equation}
  \hat{\Pi}(l) = \sum_{\Psi_{{\rm out}_2}} p(l|\Psi_{{\rm out}_2})
  |\Psi_{{\rm out}_2}\rangle\langle\Psi_{{\rm out}_2}|\,,
\label{preal}
\end{equation}
where $p(l|\Psi_{{\rm out}_2})$ is the probability of obtaining the
measurement outcome $l$ under the condition that the output reference mode 
is prepared in the state $|\Psi_{{\rm out}_2}\rangle$.
A typical example is direct photon counting with quantum
efficiency $\eta$ less than unity, 
\begin{equation}
  \hat{\Pi}(n) = \sum_{k} {k \choose n} \eta^n (1-\eta)^{k-n} 
  |k\rangle_      {2\,\,2}\langle k|\,.
\label{povm}
\end{equation}
Further, the input reference mode may also be prepared in a
mixed quantum state,
\begin{equation}
  \hat{\varrho}_{{\rm in}_2} = \sum_{\Psi_{{\rm in}_2}} p(\Psi_{{\rm in}_2}) 
  |\Psi_{{\rm in}_2}\rangle\langle\Psi_{{\rm in}_2}|\,,
\end{equation}
so that for an input quantum state 
\begin{equation}
  \hat{\varrho}_{{\rm in}} = \hat{\varrho}_{{\rm in}_1} \otimes
  \hat{\varrho}_{{\rm in}_2}
\label{in}
\end{equation}
the quantum state of the output signal mode now reads 
\begin{equation}
\label{2.9a}
  \hat{\varrho}_{{\rm out}_1} = \frac{1}{p(l)}\sum_{\Psi_{{\rm in}_2}}
  p(\Psi_{{\rm in}_2})\sum_{\Psi_{{\rm out}_2}}p(l|\Psi_{{\rm out}_2})
  \hat{Y}\hat{\varrho}_{{\rm in}_1}\hat{Y}^\dagger 
\label{generalreal}
\end{equation}
in place of (\ref{2.9}),
where $\hat{Y}$ is defined by equation (\ref{2.10}). 
In equation (\ref{2.9a}) 
\begin{equation}
  p(l) = \sum_{\Psi_{{\rm in}_2}}
  p(\Psi_{{\rm in}_2})\sum_{\Psi_{{\rm out}_2}}p(l|\Psi_{{\rm out}_2})
  {\rm Tr}_1\!\left(\hat{Y}\hat{\varrho}_{{\rm in}_1}\hat{Y}^\dagger\right).
\label{generalprob}
\end{equation}
is the probability of obtaining the measurement outcome $l$,
i.e., the probability of producing the output signal-mode
quantum state $\hat{\varrho}_{{\rm out}_1}$.

\section{Creation of Schr\"odinger-cat-like states}
\label{Sec3}

\begin{figure}
\centering\epsfig{figure=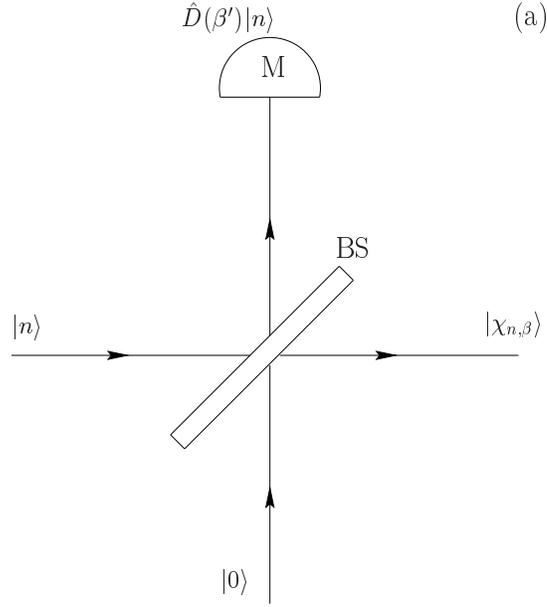,width=0.5\linewidth}

~

\vspace{1cm}
~

\centering\epsfig{figure=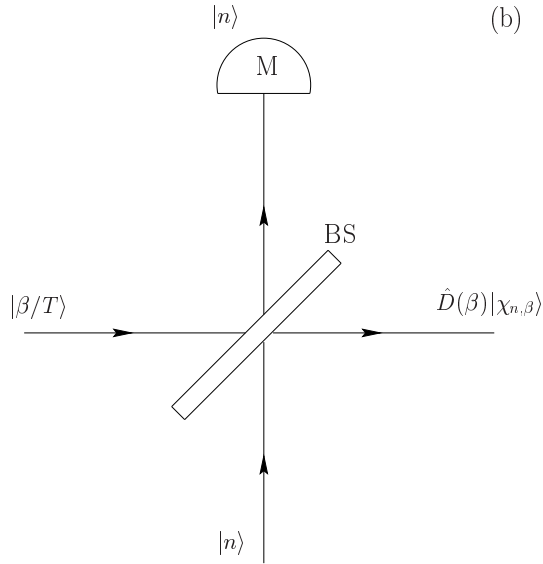,width=0.5\linewidth}
\caption{
(a) A Fock state $|n\rangle$ is fed into a balanced beam splitter. 
When the measuring instrument M projects the output reference mode
onto a displaced Fock state $\hat{D}(\beta')|n\rangle$, then
for $|\beta'|^2$ $\!=$ $\!|\beta|^2$ $\!=$ $\!n/2$ the output 
signal mode collapses to the
Schr\"odinger-cat-like state 
$|\chi_n^{n,\beta}\rangle$,
equation (\protect\ref{2.27}).\protect\\
(b) A coherent state $|\beta/T\rangle$ and a Fock state $|n\rangle$
are fed into a balanced beam splitter. When the measuring instrument M
projects the output reference mode onto the same Fock state $|n\rangle$,
then the output signal mode collapses to the
state $\hat{D}(\beta)|\chi_n^{n,\beta}\rangle$.
\label{Fig2}}
\end{figure}

\begin{figure}
\centering\epsfig{figure=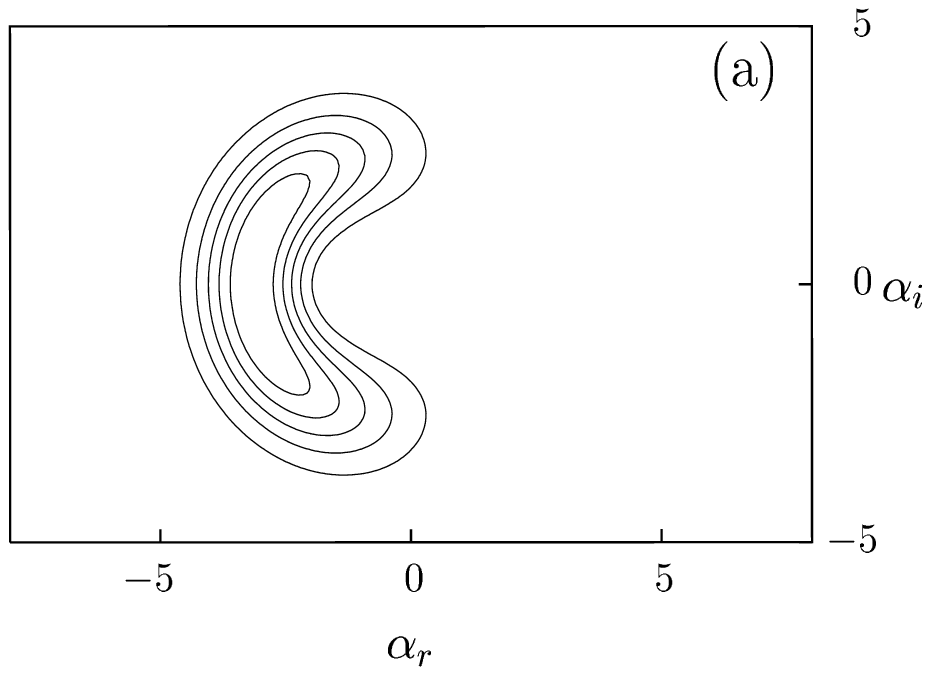,width=0.5\linewidth}
\centering\epsfig{figure=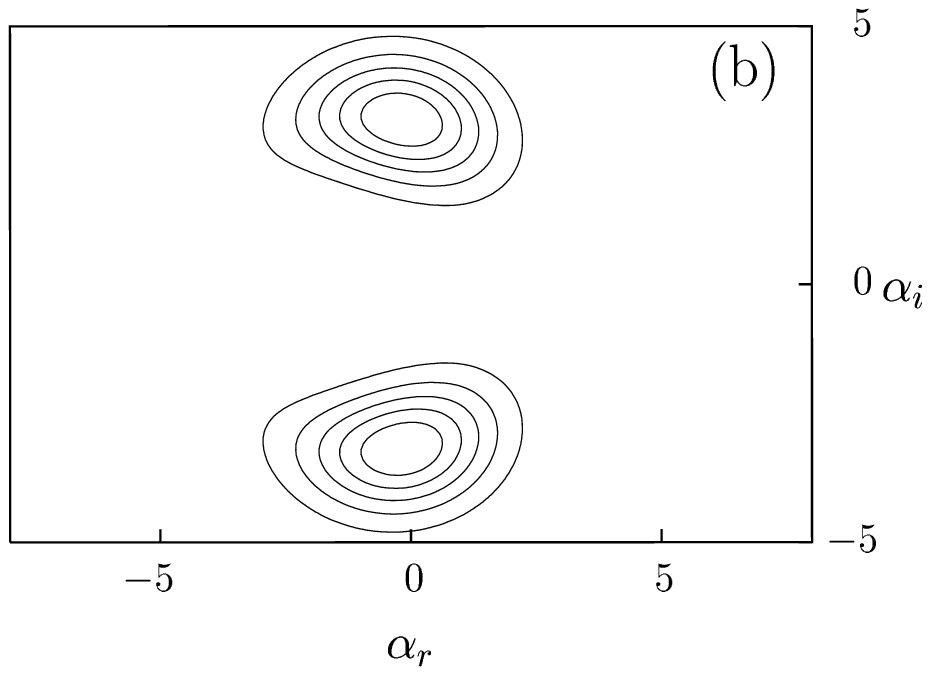,width=0.5\linewidth}
\centering\epsfig{figure=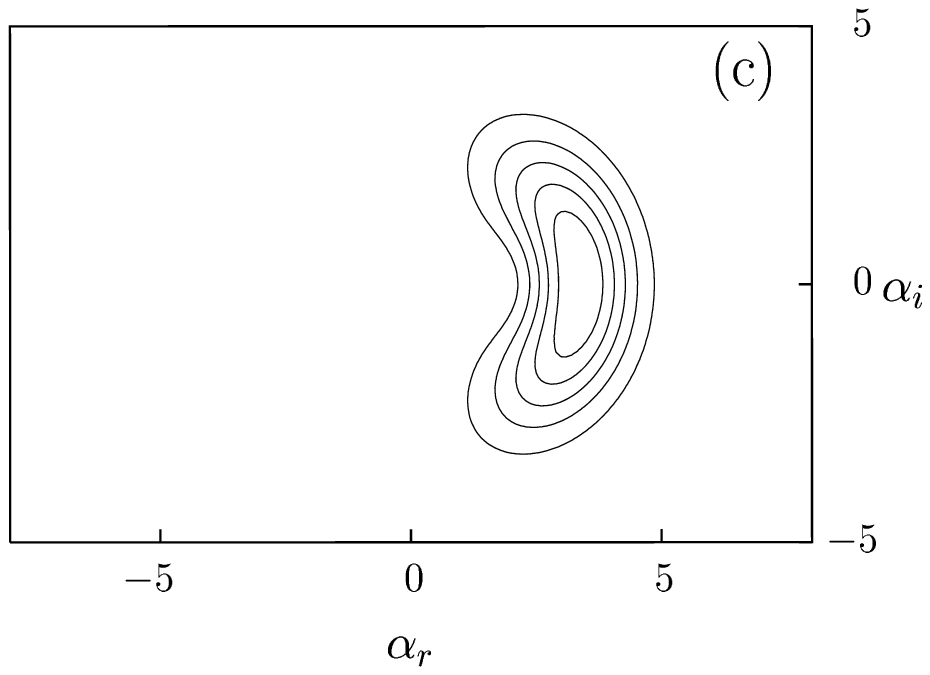,width=0.5\linewidth}
\caption{
Contour plots of the Husimi function $Q(\alpha)$, equation 
(\protect\ref{2.29}), of the state 
$|\chi_n^{n,\beta}\rangle$, equation (\protect\ref{2.27}),
for $n$ $\!=$ $\!10$ and (a) $\beta$ $\!=$ $\!\protect\sqrt{n/10}$, 
(b) $\beta$ $\!=$ $\!\protect\sqrt{n/2}$, 
and (c) $\beta$ $\!=$ $\!\protect\sqrt{2.5n}$.
\label{Fig3}}
\end{figure}

\begin{figure}
\centering\epsfig{figure=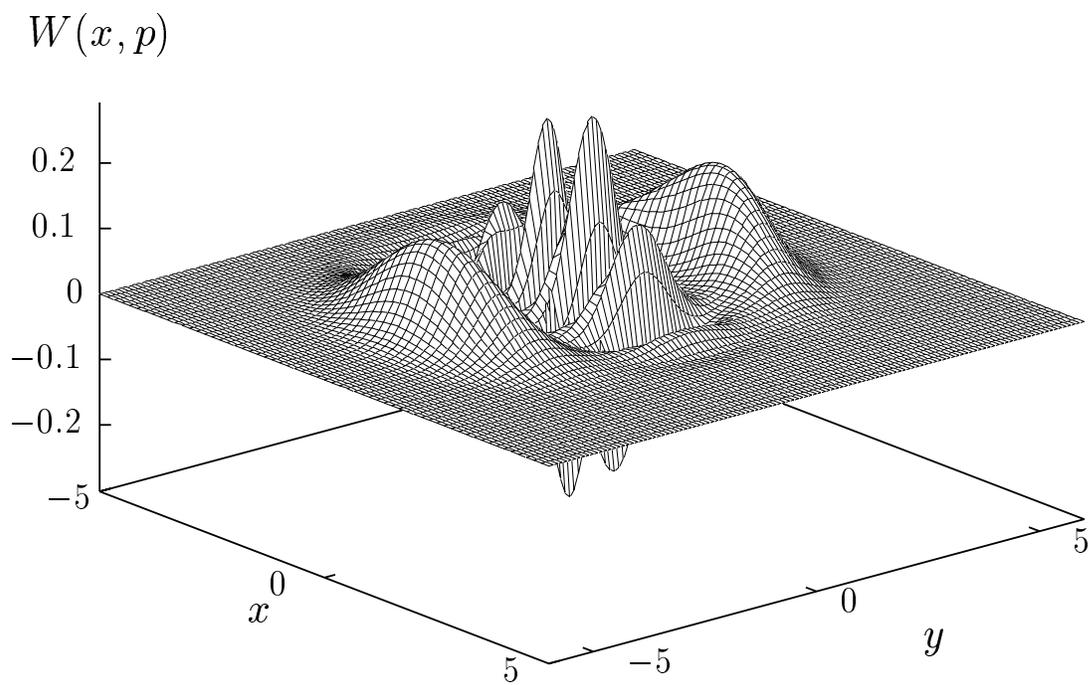,width=1\linewidth}
~

\vspace{4cm}
~
\caption{
The Wigner function $W(x,p)$, equation (\protect\ref{2.34}), of the state 
$|\chi_n^{n,\beta}\rangle$,
equation (\protect\ref{2.27}),
is shown for $n$ $\!=$ $\!10$ and $\beta$ $\!=$ $\!\protect\sqrt{n/2}$.
\label{Fig4} }
\end{figure}

\begin{figure}
\centering\epsfig{figure=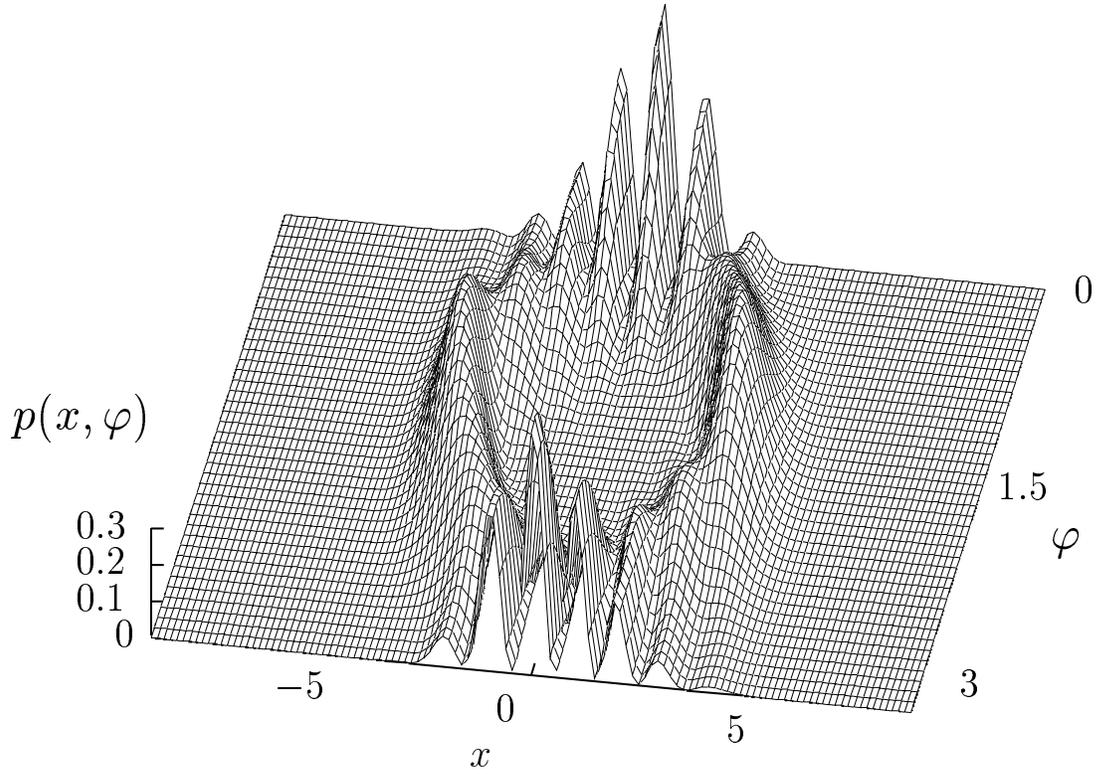,width=1\linewidth}
~

\vspace{4cm}
~
\caption{
The quadrature-component probability distributions $p(x,\varphi)$, equation
(\protect\ref{2.32a}), of the state
$|\chi_n^{n,\beta}\rangle$,
equation (\protect\ref{2.27}),
are shown for $n=10$ and $\beta=\protect\sqrt{n/2}$. 
\label{Fig5} }
\newpage
\end{figure}

\begin{figure}
\centering\epsfig{figure=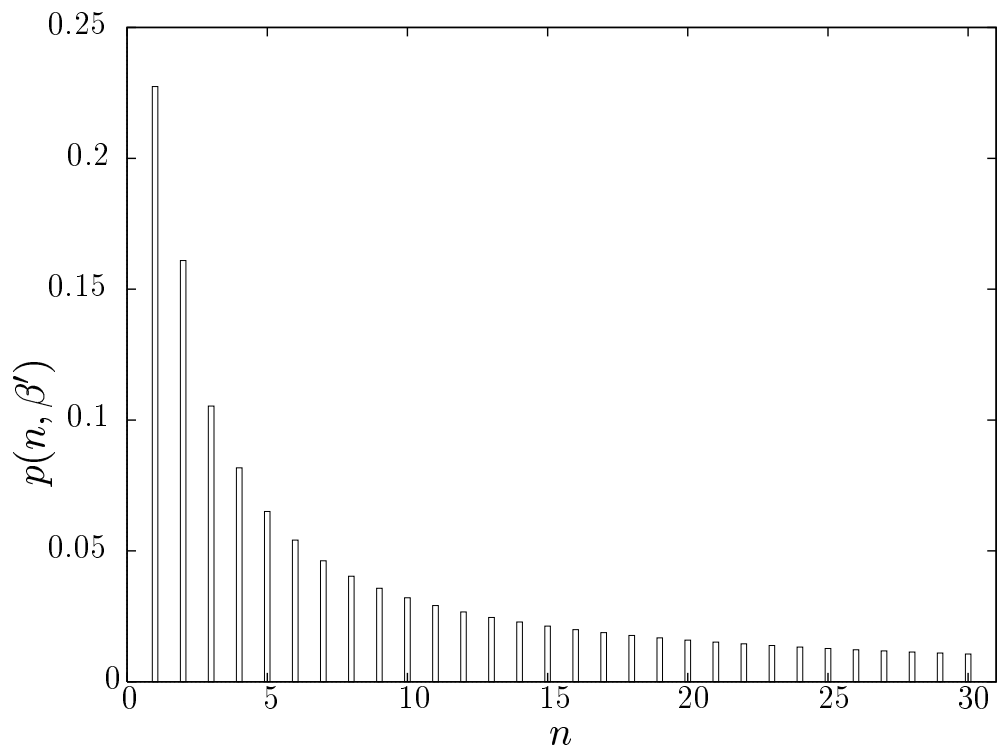,width=1\linewidth}
~

\vspace{4cm}
~
\caption{
The probability $p(n,\beta')$, equation (\protect\ref{2.28}), 
of producing the state 
$|\chi_n^{n,\beta}\rangle$,
equation (\protect\ref{2.27}),
is shown as a function of $n$ for $|\beta'|^2$ $\!=$ $\!|\beta|^2$
$\!=n/2$.
\label{Fig6} }
\end{figure}

In order to illustrate the theory, let us consider the generation
of Schr\"odinger-cat-like states. A possible way is to prepare
the input signal mode in a squeezed vacuum, combine it with
an input reference mode prepared in a Fock state (including the vacuum
state) and perform a photon-number measurement on the
output reference mode \cite{Dakna1,Dakna2}. Here we present 
two alternative schemes that are only based on Fock states and coherent
states. Let us first consider a scheme [figure \ref{Fig2}(a)] that 
uses a Fock state source
and displaced photon-number measurement,
\begin{eqnarray}
\label{2.24}
|\Psi_{{\rm in}_1}\rangle=|n\rangle ,
\end{eqnarray}
\begin{eqnarray}
\label{2.25}
|\Psi_{{\rm in}_2}\rangle=|0\rangle ,
\qquad
|\Psi_{{\rm out}_2}\rangle=\hat{D}(\beta^\prime)|n\rangle .
\end{eqnarray}
For notational convenience we omit the subscripts $1$ and $2$ 
introduced in Section \ref{Sec2} in order to distinguish between the 
channels. Application of equations (\ref{2.9}) and (\ref{2.14})
then yields ($|T|^2$ $\!=$ $\!0.5$)
\begin{eqnarray}
|\Psi_{{\rm out}_1}\rangle =: 
|\chi_n^{n,\beta}\rangle 
&=& \frac{1}{n!\sqrt{N}}
\big(\hat{a}-\beta\big)^n\big(\hat{a}^\dagger+\beta^*\big)^n|0\rangle\nonumber\\
&=& N^{-1/2}\sum_{k=0}^n {\rm L}_{n-k}^k\!\left(|\beta|^2\right)
\,\frac{(-\beta\hat{a}^\dagger)^k}{k!}\,|0\rangle\nonumber\\
&=& N^{-1/2}{\rm L}_n\!\left[\beta\hat{D}^\dagger(\beta)\hat{a}^\dagger
\hat{D}(\beta)\right]|0\rangle
\label{2.27}
\end{eqnarray}
[$\beta$ $\!=$ $\!\beta^\prime  e^{\i(\varphi_T+\varphi_R+\pi)}$;
${\rm L}_n(z)$, Laguerre polynomial; ${\rm L}_n^a(z)$,
associated Laguerre poly\-nom\-ial], where
\begin{equation}
2^{-n} e^{-|\beta|^2}\,N=2^{-n} e^{-|\beta|^2}
\sum_{k=0}^n\frac{|\beta|^{2k}}{k!}\,
{\rm L}_{n-k}^k\!\left(|\beta|^2\right)^2
= p(n,\beta'),
\label{2.28}
\end{equation}
$p(n,\beta')$ being the probability of generating the state 
$|\chi_n^{n,\beta}\rangle$ for given $\beta'$ 
[see equation (\ref{2.11})]. Equation 
(\ref{2.27}) reveals that the quantum state of output signal mode is 
obtained by applying an operator Laguerre polynomial ${\rm L}_n$ on the 
vacuum state.\footnote{Note that the states $|\chi_n^{n,\beta}\rangle$
are different from the Laguerre polynomial states introduced
in Ref. \protect\cite{Hong}. The latter can be produced in the
scheme studied in Ref. \protect\cite{Dakna3}.}
As it is seen from the in figure \ref{Fig3} plotted
contours of the Husimi function
\begin{equation}
Q(\alpha)=\frac{1}{\pi}|\langle\alpha|
\chi_n^{n,\beta}
\rangle|^2
=\frac{1}{\pi N}\left|{\rm L}_n[\beta(\alpha^\ast\!+\!\beta^\ast]\right|^2
 e^{-|\alpha|^2},
\label{2.29}
\end{equation}
the state $|\chi_n^{n,\beta}\rangle$
exhibits for $|\beta|^2$ $\!=$ $\!n/2$ two well separated peaks
in the phase space, which are approximately located at
$\pm {\rm i}\beta$ [figure \ref{Fig3}(b)].
It should be noted that the distance of the peaks increases with 
the square root of the detected photons, which is 
analogous to the behaviour of the Schr\"odinger-cat-like states 
considered in \cite{Dakna1}. The Husimi $Q$ function, which
can be directly measured, e.g., in eight-port homodyne detection
or heterodyne detection, is a rather smeared phase-space function,
because of the included vacuum noise.  
More details of the structure of the state can be inferred from the 
Wigner function
\begin{equation}
W(x,p)=\frac{1}{\pi}\int_{-\infty}^{+\infty}dy\,{\rm e}^{2{\rm i}py} 
\langle x-y,0
|\chi_n^{n,\beta}\rangle\langle\chi_n^{n,\beta}|x+y,0\rangle,
\label{2.33}
\end{equation}
where $|x,0\rangle$ is the quadrature-component state
\begin{equation}
|x,\varphi\rangle=(\pi)^{-\frac{1}{4}}
\exp\!\left(-\textstyle\frac{1}{2}x^2\right)
\sum_{k=0}^{\infty}\frac{ e^{\i k\varphi}}
{\sqrt{2^kk!}} \, {\rm H}_k(x) |k \rangle
\label{2.32}
\end{equation}
for $\varphi$ $\!=$ $\!0$ [H$_{k}(z)$, Hermite polynomial].
Using equations (\ref{2.27}) and (\ref{2.32}) and
applying standard formulas \cite{Prudnikov}, the $y$-integral in 
equation (\ref{2.33}) can be calculated to obtain
\begin{eqnarray}
\lefteqn{
\hspace*{-12ex}
W(x,p)=\frac{e^{-(x^2+p^2)}}{\pi N} 
\sum_{m=0}^{n}\Bigg\{
\sum_{k=0}^{m}{\rm L}_{n-k}^k\big(|\beta|^2\big)
{\rm L}_{n-m}^m\big(|\beta|^2\big)
{\rm L}_k^{m-k}\big(|z|^2\big)
\beta^k \big(-\beta^\ast\big)^m
\frac{z^{m-k}}{m!}
}
\nonumber \\ &&  \hspace{-5ex} 
+\!\!\sum_{k=m+1}^{n}\!\!{\rm L}_{n-k}^k\big(|\beta|^2\big)
{\rm L}_{n-m}^m\big(|\beta|^2\big)
{\rm L}_k^{k-m}\big(|z|^2\big)
\beta^k \big(-\beta^\ast\big)^m 
\frac{(-z^\ast)^{k-m}}{k!}\Bigg\}
\label{2.34}
\end{eqnarray}
[$z\!=\!\sqrt{2}(x$ $\!+$ $\!ip)$].
The Wigner function is a measurable quantity which however does not 
correspond to a POVM. Let us therefore also
consider the quadrature-component probability distributions
\begin{eqnarray}
\label{2.30}
p(x,\varphi)&=&
|\langle x,\varphi|\chi_n^{n,\beta}\rangle|^2
\nonumber\\
&=&\frac{1}{\pi^{\frac{1}{2}}N}\left|
\sum_{k=0}^n{\rm L}_{n-k}^k(|\beta|^2)
\frac{(-2^{-\frac{1}{2}}\beta^\ast  e^{\i\varphi})^k}{k!} {\rm H}_k(x)\right|^2
 e^{-x^2},
\label{2.32a}
\end{eqnarray}
which also contain all available information on the quantum state
and can be directly measured in four-port homodyne detection.
Plots of the Wigner function  
and the quadrature-component distributions  
of the state 
$|\chi_n^{n,\beta}\rangle$
for $\beta$ $\!=$ $\!\sqrt{n/2}$ are presented in figures 
\ref{Fig4} and \ref{Fig5} respectively. They clearly reveal
the typical features of Schr\"odinger-cat-like states.
The probability of producing the state for various values of
$n$ is shown in figure \ref{Fig6}.

Let us compare the scheme in figure \ref{Fig2}(a) with the somewhat
modified scheme in figure \ref{Fig2}(b). In the latter
scheme the input signal mode that is prepared in a coherent state
is mixed with an input reference mode that is prepared in a Fock state,
and a photon-number measurement is performed on the
output reference mode,
\begin{eqnarray}
\label{2.35}
|\Psi_{{\rm in}_1}\rangle&=&|\beta/T\rangle,
\end{eqnarray}
\begin{eqnarray}
\label{2.36}
|\Psi_{{\rm in}_2}\rangle=|n\rangle,
\qquad
|\Psi_{{\rm out}_2}\rangle=|n\rangle.
\end{eqnarray}
The scheme realizes the generation of a 
special Jacobi-polynomial coherent state \cite{Dakna3},
\begin{eqnarray}
|\Psi_{{\rm out}_1}\rangle &\sim& \{(\hat{a}^+)^n\hat{a}^n\}_{s=3}\,
T^{\hat{n}}\, |\beta/T\rangle
\,\sim\,{\rm P}_n^{(0,\hat{n}-n)}(0)\, |\beta\rangle
\nonumber\\
&\sim&{\rm L}_n\!\left(\beta\hat{a}^\dagger\right) |\beta\rangle
\,=\,\hat{D}(\beta)\,
|\chi_n^{n,\beta}\rangle
\label{2.38}
\end{eqnarray}
($|T|^2$ $\!=$ $\!0.5$).
Here we have used equations (\ref{2.14}), (\ref{2.15}) and
(\ref{2.17}) and the relation 
${\hat{n} \choose k}$ $\!=$ $\!(\hat{a}^\dagger)^k\hat{a}^k/k!$. 
Comparing equations (\ref{2.27}) and (\ref{2.38}), 
we see that the states produced in the schemes in figures 
\ref{Fig2}(a) and \ref{Fig2}(b) differ in a 
coherent displacement. Obviously, the coherently displaced state 
shows again for $|\beta|^2$ $\!=$ $\!n/2$ the typical features of 
a Schr\"odinger-cat-like state, the phase difference between
the component states being $\pi/2$. The probability of
producing the state is the same as in equation (\ref{2.28}).

\begin{figure}
\centering\epsfig{figure=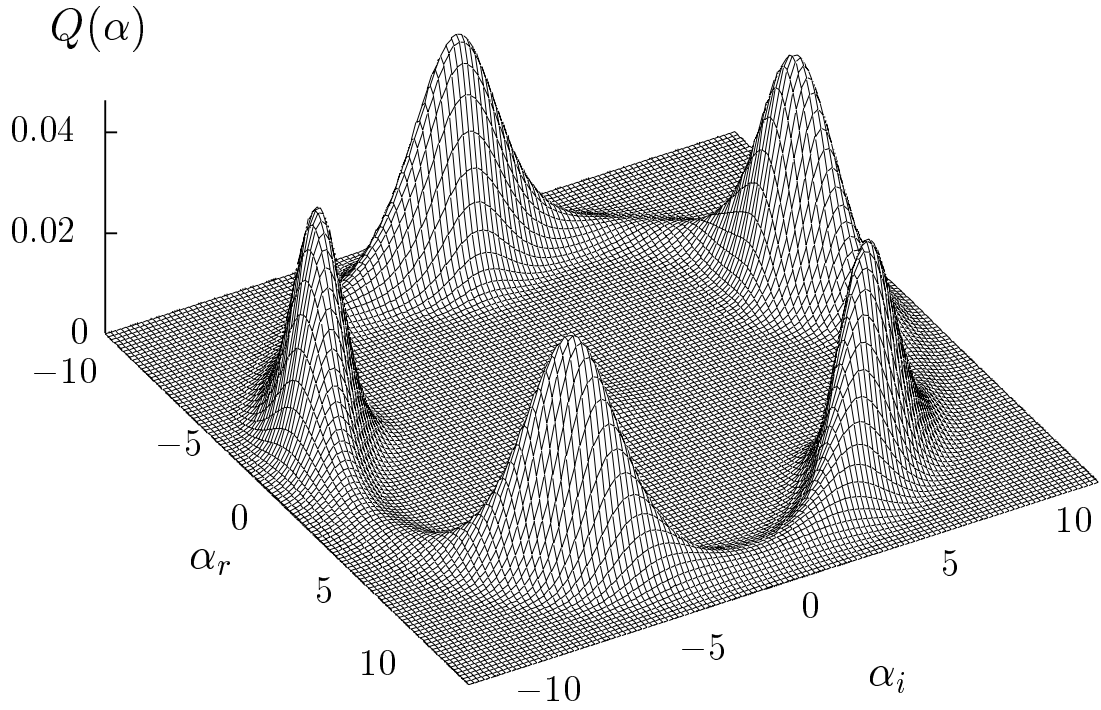,width=1\linewidth}
~

\vspace{4cm}
~
\caption{
The Husimi function $Q(\alpha)$, equation(\protect\ref{2.42a}), of 
the state $|\psi_{n,\beta}^{(k)}\rangle$, equation (\protect\ref{2.41}), 
is shown for $n$ $\!=$ $\!10$, $k$ $\!=$ $\!5$, and $\beta$ $\!=$ $\!4.2$.
\label{Fig7} }
\end{figure}

It is worth noting that replacing in equation (\ref{2.27})
$(\hat{a}$ $\!-$ $\!\beta)$ with $(\hat{a}$ $\!-$ $\!\beta)^\dagger$
yields the state
\begin{eqnarray}
\label{2.40}
|\psi_{n,\beta}^{(2)}\rangle
= \frac{1}{\sqrt{N_2}}
\left(\hat{a}^\dagger\!-\!\beta^*\right)^n
\left(\hat{a}^\dagger\!+\!\beta^*\right)^n|0\rangle
= \frac{1}{\sqrt{N_2}}\left[\left(\hat{a}^\dagger\right)^2\!-\!\left(\beta^*\right)^2\right]^n|0\rangle,
\end{eqnarray}
the properties of which are similar to those of the state
$|\chi_n^{n,\beta}\rangle$
in equation (\ref{2.27}). 
States of the type (\ref{2.40})
may be produced, e.g., by two displaced $n$-photon-additions
(for such schemes, see \cite{Dakna4}). The generalization
to $k$ displaced $n$-photon-additions is straightforward,
\begin{equation}
|\psi_{n,\beta}^{(k)}\rangle = \frac{1}{\sqrt{N_k}}
\left[\left(\hat{a}^\dagger\right)^k 
- \left(\beta^*\right)^k\right]^n|0\rangle,
\label{2.41}
\end{equation}
\begin{equation}
\label{2.42}
N_k = \sum_{j=0}^n {n \choose j}^2 |\beta|^{2k(n-j)} (kj)!.
\end{equation}
States of this type can be regarded, for appropriately chosen
$\beta$, as multiple Schr\"odinger-cat-like states \cite{Janszky},
as it can be seen from figure \ref{Fig7}, in which the
Husimi function 
\begin{equation}
\label{2.42a}
Q(\alpha)=\frac{1}{\pi N_k}
\left|\alpha^k-\beta^k\right|^{2n} e^{-|\alpha|^2}
\end{equation}
of such a state is plotted.

\section{Conclusions}
\label{Sec5}

In this paper we have studied conditional quantum-state generation 
at a beam splitter, regarding the apparatus as an effective two-port
device, which is described by a non-unitary beam splitter 
operator $\hat{Y}$ that acts in the Hilbert space of the signal mode only.
We have presented the $\hat{Y}$ operator for arbitrary
quantum states $|\Psi_{{\rm in}_2}\rangle$ and $|\Psi_{{\rm out}_2}\rangle$
of the reference mode as an $s$-ordered operator product of those operators
which generate $|\Psi_{{\rm in}_2}\rangle$ and  $\langle\Psi_{{\rm out}_2}|$
from the vacuum, $s$ being entirely determined by the absolute value of 
the reflectance (or transmittance) of the beam splitter. 
We have further 
given a generalization to the case of mixed input states and non-perfect 
measurement. 

The formalism is a generalization of previously obtained results, which
are easily recovered by appropriate specification of the non-unitary 
transformation operator. Given the quantum state of the input
signal, the conditional quantum state of the output signal can be 
calculated in a very straightforward way for arbitrary input reference states
and arbitrary measurement-assistant projection. The probability of 
producing the conditional quantum state of the output 
signal is simply given by its norm. In order to illustrate the theory,
we have considered two schemes for generating Schr\"odinger-cat-like 
states and addressed the problem of generating multiple 
Schr\"odinger-cat-like states. Finally, it should be pointed out that the 
formalism developed is also suitable for studying state manipulation and 
state engineering via conditional measurement at multiports that are 
built up by beam splitters, such as chains of beam splitters.

\begin{appendix}
\renewcommand{\thesection}{Appendix \Alph{section}}
\section{Derivation of Equation (\protect\ref{2.14})}
\label{app1}
\setcounter{equation}{0}
\renewcommand{\theequation}{\Alph{section}\arabic{equation}}
  From equations (\ref{2.10}) and (\ref{2.12}) the
operator $\hat{Y}$ is defined by
\begin{equation}
\label{ska}
\hat{Y} = \,_2\langle n|\hat{D}_2^\dagger(\beta)\hat{U}
\hat{D}_2(\alpha)|m\rangle_2\,,
\end{equation}
where $\hat{U}$, equation (\ref{2.2}), can be rewritten as \cite{Dakna2}
\begin{eqnarray}
\label{a1}
\hat{U} = e^{\i(\varphi_T\!+\! \varphi_R)\hat{L}_3} e^{2\i\vartheta\hat{L}_2}
e^{\i(\varphi_T - \varphi_R)\hat{L}_3} 
=T^{\hat{n}_1} e^{-R^\ast\hat{a}^\dagger_2\hat{a}_1} e^{R\hat{a}^\dagger_1
\hat{a}_2}T^{-\hat{n}_2}\,.
\end{eqnarray}
We insert equation (\ref{a1}) into equation (\ref{ska}) 
and calculate the channel-2 matrix element as
\begin{equation}
\hat{Y}=\frac{1}{\sqrt{m!n!}}\;_2\langle\beta|(\hat{a}_2
-\beta)^nT^{\hat{n}_1} e^{-R^\ast\hat{a}^\dagger_2\hat{a}_1} e^{R\hat{a}^\dagger_1\hat{a}_2}
T^{-\hat{n}_2}(\hat{a}_2^\dagger-\alpha^\ast)^m|\alpha\rangle_2,
\end{equation}
where we have used the relation 
\begin{equation}
\hat{D}^{-1}(\alpha)f(\hat{a},\hat{a}^\dagger)\hat{D}(\alpha)=
f(\hat{a}\!+\! \alpha,\hat{a}^\dagger\!+\! \alpha^\ast)
\label{a2}
\end{equation}
and $|n\rangle\!=\!1/\sqrt{n!}(\hat a^\dagger)^n|0\rangle$.
Applying the relations
\begin{eqnarray}
\nonumber
e^{\varphi\hat{n}}f(\hat{a})=f(e^{-\varphi}\hat{a})e^{\varphi\hat{n}},
\qquad
e^{\varphi\hat{n}}f(\hat{a}^\dagger)=f(e^{\varphi}\hat{a}^\dagger)
e^{\varphi\hat{n}},
\\
 e^{\varphi\hat{a}^\dagger}f(\hat{a})
=f(\hat{a}\!-\!\varphi)e^{\varphi\hat{a}^\dagger},
\qquad
e^{\varphi\hat{a}}f(\hat{a}^\dagger)
=f(\hat{a}^\dagger\!+\!\varphi)e^{\varphi\hat{a}},
\label{Rel}
\end{eqnarray}
after straightforward algebra we derive
\begin{eqnarray}
\lefteqn{
\hat{Y}=\frac{T^{\hat{n}_1} e^{-R^\ast\beta^\ast\hat{a}_1}}{\sqrt{m!n!}}
\sum_{k=0}^{m}\sum_{l=0}^{n}{m \choose k}{n \choose l}T^{-k}
\left(-\beta\!-\!R^\ast\hat{a}_1\right)^{n-l}
}
\nonumber\\&&\hspace{10ex}\times\;
\left[-\alpha^\ast\!+\!\frac{R}{T}
\,\hat{a}_1^+\right]^{m-k}\!_2\langle\beta|\hat{a}_2^l
(\hat{a}_2^\dagger)^k
T^{-\hat{n}_2}|\alpha\rangle_2\,
e^{(R/T)\alpha\hat{a}^\dagger_1}.
\end{eqnarray}
Using the ordering formula \cite{Cahill}  
\begin{eqnarray}
\hat{a}^{m}(\hat{a}^\dagger)^n=\sum_{l=0}^{\min\{m,n\}}{m\choose l}
\frac{n!}{(n-l)!}
(\hat{a}^\dagger)^{n-l}\hat{a}^{m-l}
\label{ord}
\end{eqnarray}
together with equation (\ref{Rel}) and the
relation
\begin{equation}
  T^{\hat{n}}|\alpha\rangle =
  e^{-|\alpha|^2(1-|T|^2)/2}|T\alpha\rangle,
\end{equation}
we have
\begin{eqnarray}
\lefteqn{
\hspace*{-12ex}
\hat{Y}=\frac{T^{\hat{n}_1}}{\sqrt{m!n!}} 
\,e^{\alpha\beta^\ast/T} e^{-(|\alpha|^2+|\beta|^2)/2}
\sum_{k=0}^{m}\sum_{l=0}^{n}\sum_{j=0}^{\min\{k,l\}}
{m \choose k}{n \choose l}j!{k \choose j}{l \choose j}(\beta^*)^{k-j}
}
\nonumber\\&&\hspace{-4ex}\times
\alpha^{l-j}\,T^{j-k-l}\,
e^{-R^*\beta^*\hat{a}_1}
\left(-\beta\!-\!R^\ast\hat{a}_1\right)^{n-l}
\left(-\alpha^\ast\!+\!\frac{R}{T}
\,\hat{a}_1^\dagger\right)^{m-k}
e^{(R/T)\alpha\hat{a}^\dagger_1}.
\end{eqnarray}
Changing the summation indices, some of the finite summations 
may be performed to obtain
\begin{eqnarray}
\lefteqn{
\hspace*{-8ex}
\hat{Y}=
\frac{T^{\hat{n}_1-m}}{\sqrt{m!n!}} 
\,e^{\alpha\beta^\ast/T} e^{-(|\alpha|^2+|\beta|^2)/2}
\sum_{k=0}^{\min\{m,n\}}{m \choose k}{n \choose k}k!
}   
\nonumber\\&&\hspace{-4ex}\times\, 
e^{-R^*\beta^*\hat{a}_1}
\left(-R^*\hat{a}_1\!+\!\frac{\alpha}{T}\!-\!\beta\right)^{n-k}
\left(R\hat{a}_1^\dagger\!-\!T\alpha^\ast\!+\!\beta^\ast\right)^{m-k}
e^{(R/T)\alpha\hat{a}^\dagger_1}.
\end{eqnarray} 
Applying again the relations (\ref{a2}) and (\ref{Rel}), after some 
calculation we obtain
\begin{eqnarray}
\lefteqn{
\hspace*{-8ex}
\hat{Y}=
\frac{T^{-m}}{\sqrt{m!n!}}
\,\hat{D}_1\!\left(\frac{\alpha\!-\!T\beta}{R^*}\right)    
\Bigg\{
\sum_{k=0}^{\min\{m,n\}}k!{m \choose k}{n \choose k}
\left[\frac{-1\!-\!(2|R|^{-2}-1)}{2}\right]^k
}
\nonumber\\&&\hspace{-4ex}\times\,
(-|R|^2)^k\left(-\frac{R^*}{T}\right)^{n-k}(RT)^{m-k}    
\hat{a}_1^{n-k}(\hat{a}_1^\dagger)^{m-k}T^{\hat{n}_1}
\Bigg\}
\hat{D}_1\!\left(\frac{\beta\!-\!T^\ast\alpha}{R^\ast}\right).
\end{eqnarray}
Making use of the standard ordering formula \cite{Cahill}
\begin{equation}
   \left\{(\hat{a}^\dagger)^m\hat{a}^n\right\}_s 
   = \sum_{k=0}^{\min\{m,n\}}k!{m \choose k}{n \choose k}
   \left(\frac{t-s}{2}\right)^k
   \left\{(\hat{a}^\dagger)^{m-k}\hat{a}^{n-k}\right\}_t 
\label{ordering}
\end{equation}
for $t$ $\!=$ $\!-1$, we eventually arrive at equation (\ref{2.14})
(for more details, see \cite{Clausen1}).
\end{appendix}

\section*{References}
\bibliographystyle{unsrt}

\newpage
\newpage
\newpage
\newpage
\newpage
\newpage
\end{document}